\documentclass[sigconf]{acmart}
\AtBeginDocument{%
  \providecommand\BibTeX{{%
    \normalfont B\kern-0.5em{\scshape i\kern-0.25em b}\kern-0.8em\TeX}}}

\copyrightyear{2021}
\acmYear{2021}
\setcopyright{acmlicensed}\acmConference[AIES '21]{Proceedings of the 2021 AAAI/ACM Conference on AI, Ethics, and Society}{May 19--21, 2021}{Virtual Event, USA}
\acmBooktitle{Proceedings of the 2021 AAAI/ACM Conference on AI, Ethics, and Society (AIES '21), May 19--21, 2021, Virtual Event, USA}
\acmPrice{15.00}
\acmDOI{10.1145/3461702.3462582}
\acmISBN{978-1-4503-8473-5/21/05}

\begin{document}

\title{Measuring Lay Reactions to Personal Data Markets}
\fancyhead{}
\author{Aileen Nielsen}
\email{aileen.nielsen@gess.ethz.ch}
\affiliation{%
  \institution{ETH Zurich, Center for Law and Economics}
  \streetaddress{Haldeneggsteig 4}
  \city{Zurich}
  \country{Switzerland}
  \postcode{CH-8092}
}

\renewcommand{\shortauthors}{Nielsen}

\begin{abstract}
The recording, aggregation, and exchange of personal data is necessary to the development of socially-relevant      machine learning applications. However, anecdotal and survey evidence show that ordinary people feel discontent and even anger regarding data collection practices that are currently typical and legal. This suggests that personal data markets in their current form do not adhere to the norms applied by ordinary people. The present study experimentally probes whether market transactions in a typical online scenario are accepted when evaluated by lay people. The results show that a high percentage of study participants refused to participate in a data pricing exercise, even in a commercial context where market rules would typically be expected to apply. For those participants who did price the data, the median price was an order of magnitude higher than the market price. These results call into question the notice and consent market paradigm that is used by technology firms and government regulators when evaluating data flows. The results also point to a conceptual mismatch between cultural and legal expectations regarding the use of personal data.
\end{abstract}


\maketitle

\section{Introduction}
The collection and sale of personal data is an important and common activity, both in the technology sector and also in more traditional domains. But, the digital collection and sale of personal data emerged as a market activity through technological fiat rather than through cultural evolution or consensus-building \cite{Zuboff19}.\footnote{\cite{Zuboff19} has depicted the emergence of personal data markets and other elements of surveillance capitalism as akin to the European colonial invasions of the Americas, in which  distinct cultural entities, Native Americans versus Europeans, had drastically different conceptions of what their exchanges meant as well as widely disparate levels of technical capacity to  shape reality to their desired outcomes.}  Thus technology firms have evolved market infrastructures for personal data, but cultural norms may not have incorporated or accepted such market-like categories and logic. 

Rather, concomitant with the emergence of markets for personal data has been a sense of apathy or defeat by users of common digital technologies \cite{Turow15}. Consumers have developed a sense that resistance to personal data collection and sales is futile, and they have good reason to feel this way. “Do not track” and other privacy requests by consumers have been routinely disregarded \cite{Hill18}, \cite{Fleishman19}, \cite{Zuboff19}. Tech companies and other entities who collect data and build machine learning models have likewise routinely violated their own privacy policies \cite{Fair19}. Perceived and actual transgressions of privacy norms have also occurred in the case of data collection for academic research \cite{Crawford19}, \cite{Rea19}. The resulting widespread discontent  has led to consistent calls for legal reform \cite{McCabe19}, as much behavior found offensive by ordinary people is nonetheless entirely legal.\footnote{Note that not all behaviors described in this paragraph are legal under current U.S. law. The main point, however, is that even the legal behaviors anger ordinary people.}

Yet, U.S. federal lawmaking has been dormant with respect to the proliferating sales of personal data and the routine commercial surveillance that characterize current lived digital experiences. In particular, U.S. privacy law remains grounded in a market metaphor accepted from the technical fait accompli of technology firms, without any indication that data subjects - that is ordinary people -      agree with the use of commercial markets and market transactions to distribute personal data.\footnote{Nations in the European Union are considered to be more protective of personal data privacy, but their data protection laws also recognize and facilitate personal data markets premised on notice and consent. Thus the results of this paper apply more widely than the U.S., but an international analysis of privacy law is beyond the scope of this work.}

The presumption of market appropriateness implicit in laws that facilitate personal data markets is strongly undermined by the  privacy attitudes of lay people. Anecdotal data and recent experimental results alike (described in the next section) suggest that monetary indicators and market trading for privacy do not provide reliable indicators of the value of privacy to ordinary people. The present study extends this recent line of work by exploring the question of whether lay responses to market transactions for personal data are consistent with expected market behaviors. The study explores the possibility that the existence and legality of personal data markets may not coincide with acceptance by lay people of market transactions or market-like behaviors. 

This experimental study presents evidence that data privacy is mismatched with market mechanisms, and that market allocation of personal data fails to respect the cultural and contextual expectations of ordinary people \cite{Nissenbaum10}.
Specifically, this study probes the degree to which ordinary people are willing to participate in a market-like exercise of pricing personal data when they have the option not to do so. The results show that many ordinary people reject market-like rules for the allocation of personal data despite this being the dominant paradigm in privacy law and commercial practice.      
\section{Related Work}

It has long been recognized that stated and revealed preference are at odds when it comes to valuing privacy, a phenomenon known as the “privacy paradox”. Specifically, people state that privacy is very important to them but often fail to behave accordingly either in laboratory or field experiments \cite{Acquisti15}. Such results have been obtained in many and varied circumstances, including when privacy is made highly salient. 

The privacy paradox did not emerge with the most recent advances in digital technologies but has been documented for decades \cite{Spiekermann01}. Nonetheless it remains important to legal and ethical considerations of socially relevant AI technologies, many of which  rely on the availability of personal data, that is, data about people.

The question of how to value privacy, and whether markets are an appropriate mechanism to do so, has a long intellectual history. \cite{Hirshleifer80}  questioned the assumption that a market-like mechanism populated by rational actors would result in desirable social outcomes. More recently, \cite{Taylor04} has shown that the ability to make rational data control or sharing decisions is important to welfare outcomes in privacy markets. Thus the question of whether people can behave as rational economic agents with respect to their privacy is crucial to the question of whether markets are an appropriate venue for privacy. Yet, as \cite{Lee18} noted in a working paper, existing theoretical or behavioral scholarship on privacy leaves undiscussed the existence and implications of rational privacy preferences or choices.

\cite{Lee18} were the first to address explicitly the existence of rational privacy preferences. They did so with an experiment in which personal data points (body fat percentage and IQ) were measured. Research subjects were then presented with a variety of privacy options in which they traded off which of the two collected kinds of information might be shared to various audience sizes.\footnote{\cite{Lee18}'s experiment had a surprisingly high rate of attrition, suggesting that some participants had a strong reaction against the subject or mechanism of the experiment. Despite the sunk costs of time and effort to travel to the research location,  4 \% of research participants declined to participate fully in the experiment (and also despite monetary incentives to do so). This is relevant given this study’s results, to be discussed infra, that a high rate of participants refused to participate in the pricing exercise.}   \cite{Lee18} thus developed multiple data privacy trade-off points for an individual and so could assess whether there was a consistent pattern of tradeoff decisions. 

\cite{Lee18} found that a majority (63 \%) of participants did evince consistent privacy preferences when trading off various privacy scenarios against each other. But, a substantial minority (37 \%) deviated from rational orderings. Use of monetary metrics for privacy trade-offs decreased the rate of logically-consistent choices as compared to when participants made direct trade-offs of privacy options. Use of monetary metrics resulted in a majority (53 \%) of participants deviating from rational orderings.      

\cite{Lee18}'s results evince a mixed answer to the question of whether individuals have rational privacy preferences. Their results suggest that in a privacy-salient laboratory experiment, a majority of participants make rational choices in a highly unrealistic scenario in which different privacy scenarios are traded against one another. On the other hand, only a minority maintain rational consistency once privacy is traded for money.\footnote{The reduced adherence to rational orderings with the use of monetary metrics is consistent with findings from the taboo trade-offs literature \cite{Tetlock00}. The taboo trade-offs literature shows that individuals demonstrate exaggerated bounded rationality effects when making culturally taboo trade-offs.}  

Also notable, \cite{Lee18}'s results do not establish whether individuals do evince rational responses in privacy markets but merely that, for a majority of participants in a highly stylized laboratory experiment, rational trade-offs are possible. This suggests the need for an experiment that looks to some behavior indicating what people want to do or naturally do rather than what they are able to do when incentivized in a lab.\footnote{Analogously, it is well known that people, even experts, are rather poor at Bayesian thinking when not working through the formulas formally. Yet, this does not negate the fact that many statistics students are able to come to correct results with Bayesian methodologies on statistics exams. Thus it is a matter of what people tend to do versus what they can do given strong incentives.} 

Related to the question of whether people want to engage in market-like behaviors regarding personal data is experimental evidence from \cite{Svirsky19}. \cite{Svirsky19} finds that people are more willing to sell personal information when allowed to remain strategically ignorant of the associated privacy policy entailed by the sale. This occurs despite the fact that the privacy policies were short ones that would be nearly costless to read, suggesting that participants derived value directly from not knowing the privacy policy rather than from avoiding the work associated with reading a lengthy privacy policy. \cite{Svirsky19}'s results suggests a distaste for mixing markets and privacy explicitly.

In another recent work, a large-scale survey of American adults suggested a “superendowment effect” related to pricing personal data. Survey respondents evidenced a large disparity for the monthly rate they would be willing to pay to preserve the privacy of their data (the survey’s measure of willingness to pay) as compared to the access fee they would charge for their data (willingness to accept), with median values of \$5 and \$80 respectively \cite{Winegar19}. The authors concluded from this finding that “little or no attention” should be given to either willingness to pay or willingness to accept measures, a conclusion consistent with the hypothesis that ordinary people reject market metrics for personal data. More broadly, their results suggest that experiments and real world behavioral data alike may produce “paradoxical” results because markets and monetary measures are      inappropriate for personal data privacy.

Experiments have thus amassed a wide and varied set of behavioral facts suggesting that many people do not engage in consistent behaviors expected from rational economic agents in commercial privacy markets. This study poses an orthogonal question. Do ordinary people find such markets acceptable? Or in other words, do such marketplaces make sense in the conceptual categorizations people have for personal data? Concretely, the research question is operationalized by whether research participants will provide a price for personal data sales when they are given the option not to do so, among other similar indicators of privacy market acceptability.

\section{Experiment Overview}

The experiment seeks to determine whether judgments about personal data sales evince known patterns for market goods. If personal data are market goods, pricing behavior should follow patterns already documented for market goods in previous experiments, such as that of \cite{McGraw05}. On the other hand, if personal data do not fit well into market categories, the measured pattern of pricing behavior will deviate from those expected for a typical market good.

The structure of the experiment is designed to mirror that of \cite{McGraw05}. They studied how relational framing influenced the judged appropriateness of a graduate student, John, selling a pen. In the original study, research participants read how John came into possession of the pen either through receiving it as a gift in thanks from another member of the graduate research laboratory for help (the non-market condition) or through buying it from a colleague who sources wholesale pens and resells them for a small profit (the market condition).\footnote{\cite{McGraw05} used four kinds of relational framing drawn from \cite{Fiske91}, one of which was a market framing. Here the framing is simplified to a market framing and a non-market framing, with the latter corresponding to equality matching in \cite{McGraw05}. }

\cite{McGraw05}'s results showed that the appropriateness of selling a typical market good was influenced by its relational framing. Distress regarding the transaction was higher for the non-market condition than for the market condition. All participants provided a price in the case of the market framing but fewer than 80 \% of participants provided a price in the non-market framing. In all cases, the provided price was of the same order of magnitude as the market price, although participants asked a higher price in the case of the non-market framing. This last result suggests that participants used pricing for an expressive purpose, similar to the results and discussion of \cite{Winegar19}.

If personal data are like market objects, the current study should produce market-like behavioral patterns.  Specifically, a relational framing of a market category should make personal data sales acceptable, as indicated by high willingness to price the data and data prices that are close to the market price. On the other hand, if personal data are not like typical market goods, one would expect to see pricing behavior that does not conform to market expectations even in the case of market framing. In such a case, participants should refuse to price at non-zero rates even in the case of a market framing. Also, the pricing of the data should deviate substantially from the market rate. Such behaviors would manifest the distressed behavior and expressive pricing \cite{McGraw05} identified in cases where a transaction was judged to be inappropriate. 

\section{Design}

This study uses a vignette methodology. In the vignette,\footnote{The full vignette and experimental screen flow is available in the supplementary materials, available at https://osf.io/de5hc/.}  John has a website that is a messaging platform either for college alumni networking or for buying and selling auto parts. John’s website is highly successful, and he is approached by a data broker seeking to purchase the website’s data. John needs to decide whether to sell the data and if so how much to charge. The proposed transaction is a typical and legal one.

The study had a 2 x 3 between-subjects factorial design, varying the relational framing of the data as represented by the website where the data was collected (a college alumni messaging platform or a messaging platform to buy and sell auto parts) and the form of information provided about data market price (no information, explicit market information, or veiled information). 

The relational framing of the data was varied because past results indicate that a market framing should make a subsequent sale of an object acceptable, as indicated by all participants agreeing to provide a price. 

The pricing information was varied for two reasons. First, the no information and explicit information treatments provided a way to control for background belief or assumption by participants regarding the likely market price.  Second, the veiled information treatment provided a way to gauge participant interest in knowing the price by having the option to know or not to know the price.

The first four questions mirrored those \cite{McGraw05} used to create a composite distress measure, with four Likert-scale judgments about the acceptability of the transaction. In this way, participants’ overall affective reaction to the transaction was measured. 

Next, two questions were presented regarding the data sale in quantitative market terms. First, participants were asked how much John should charge per data point. Second, participants were asked what percentage of the proceeds John should share with the website users.     
\section{Procedure}

Participants were randomly assigned to one of the six conditions reflecting the factorial design and read the appropriate vignette. Participants then responded to four Likert scale questions (on a scale of 7) regarding the judged appropriateness of John selling the data.      Participants answered a question regarding how much John should ask per data point if he sells the data. Potential responses were provided in multiple choice format and ranged in successive orders of magnitude from \$.001 to \$1,000 per data point.\footnote{The market price of \$.01 that was chosen for the data on a per-data-point basis was set based on media coverage of market value for similar data, likely representing an upper bound rather than actual price for such data. The decision to round up (as compared to published market prices of \$.006 or \$.007) was taken to simplify the numerical data for participants and avoid confusion regarding the number of zeros in the price.}  Participants also had the option to refuse to provide a price and indicate that the sale should not take place.

The pricing treatment was applied in the pricing question. There were three pricing treatment options. Coupled with the pricing question was either no information about the typical market sale, explicit information (“Such information usually sells for around \$.01 per user, that is, about a cent per user.”), or the option to click a toggle to show the information about typical market price (revealing the same text as in the explicit information treatment). The pricing treatment applied only after the distress indicators regarding the transaction were already collected, and so analysis for pricing includes pricing treatments but analysis for distress does not. 

Following the pricing question, participants answered a question regarding what portion of the proceeds John should pass back to his customers, with multiple choice options ranging from 0\% up to 95\%. As with the pricing question, participants could also refuse to choose a percentage and instead indicate that a sale should not take place. 

Finally, participants answered two comprehension checks regarding the manipulations as well as a series of demographic questions./footnote{Demographic data are reported in the online supplementary materials at https://osf.io/de5hc/.} 

Following preregistration of the experiment (https://osf.io/de5hc/), data was collected in January 2021 on a sample of 346 U.S. adults via the Prolific online platform. Of those participants, 295 passed the two comprehension checks and were included in the data analysis. All data selection and analyses were pre-registered unless identified as post hoc.

\section{Results}

\subsection{Distress regarding the transaction}

A distress indicator was calculated as the arithmetic mean of the four Likert scale outrage indicators. There was no difference between the relational framings as measured by the mean distress about the proposed data sale.

\begin{figure}[h]
  \centering
  \includegraphics[width=\linewidth]{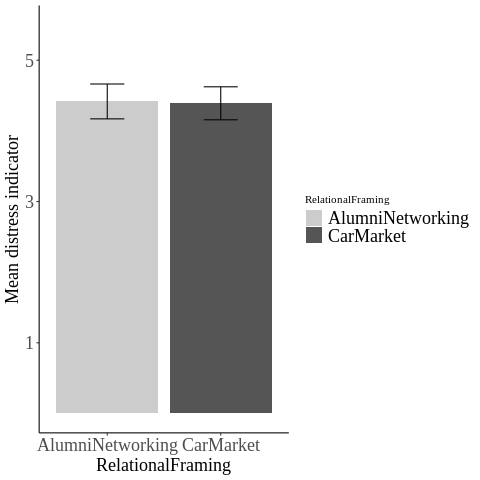}
  \caption{Mean composite distress indicator for each relational framing. Error bars are +/- two standard errors.}
  \Description{Mean distress.}
\end{figure}

These results were consistent with a pre-registered equivalence hypothesis with a large effect size (Cohen’s d = 1.0, t(293) = -7.622, p < .0001, 90\% confidence interval (-.022, .073)).  The results suggest that distress regarding the proposed data sale does not result from the relational framing of the data but rather from the nature of the data itself.

\subsection{Interest in seeing the price}

More than 85\% of participants who were offered the option to see the market price information opted to do so (the veiled treatment condition). Yet a large portion (40\%) of the participants who opted to see the price went on to refuse to provide a price for the data. These results suggest that refusal to price personal data need not be related to ignorance of or disinterest in the market price.

\subsection{Refusal to price}

More than 40\% of participants refused to price regardless of treatment condition (p < .01\footnote{Full summary statistics reported in the supplementary materials at https://osf.io/de5hc/.}).  These results diverge significantly from the expected pattern for a typical market good, for which the refusal-to-price rate would presumptively approach 0\%.\footnote{For example, in \cite{McGraw05}'s results, there was a 0\% refusal to price in the market framing condition.} 

\begin{figure}[h]
  \centering
  \includegraphics[width=\linewidth]{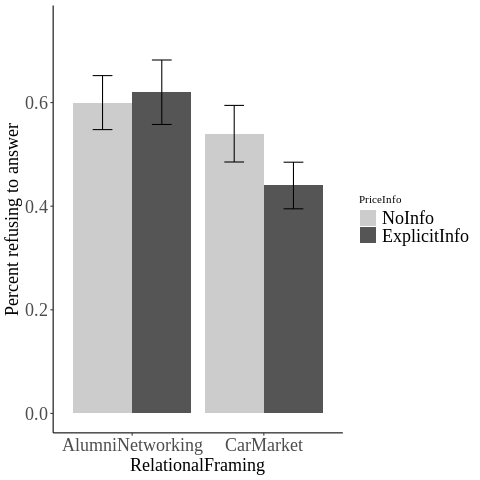}
  \caption{Percent refusing to provide a price. Error bars are +/- two standard errors.  
}
  \Description{Refusal to price.}
\end{figure}

While Figure 2 suggests a possible difference in the refusal to price rates according to relational framing, this is not borne out by post hoc statistical testing.\footnote{There is not a statistically significant difference between the relational framing for either the no information condition (W = 1130, p = .56) or for the explicit information condition (W = 991.5, p = .08).}  A larger sample size might render detectable some effect of relational framing. Importantly, however, any such effect is clearly smaller than the effect of personal data itself as a good that participants chose not to price at high rates. Also, the results in Figure 2 suggest that any such effect would go in the expected direction that a market framing would make the sale acceptable to a larger percentage of people as compared to the non-market relational framing.     

\subsection{Distribution of provided prices}

Of the prices provided, the median price was significantly above the market price (\$.01) in three of the four examined cases (p < .05\footnote{Full summary statistics reported in the supplementary materials at https://osf.io/de5hc/.}), as shown in Figure 3. \footnote{Again, these results are distinct from \cite{McGraw05}'s results, in which prices were at or slightly above market price for all treatments.}    

\begin{figure}[h]
  \centering
  \includegraphics[width=\linewidth]{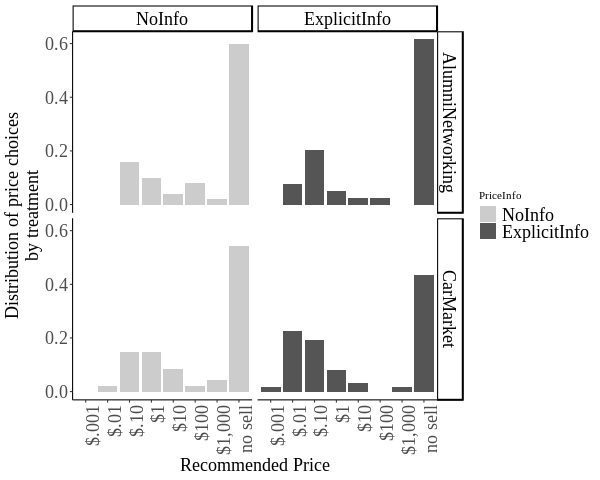}
  \caption{Distribution of prices, including refusal to price because data should not be sold.  
}
  \Description{Refusal to price.}
\end{figure}

The full distribution of pricing is plotted in Figure 3. There may be differences in distribution between treatments, as influenced by price information and relational framing, but the current sample size combined with the high rate of refusal to price suggests that applying post hoc testing on the distribution is unlikely to detect any possible effect. This is left to future work with a larger sample.

\subsection{Distribution of recommended revenue-sharing percentages}

After selecting a price for the data, participants were given the option to recommend the percentage of the proceeds the website owner should share with the data subjects. This scenario might have been expected to make the data sale more acceptable, but the rate of refusing to price was the same as before participants were made aware of the option to share proceeds. A post hoc analysis indicated that the rate of refusing the sale even with the possibility of sharing was statistically equivalent to the rate of refusing the sale when no such possibility was mentioned (TOST analysis, Cohen’s d = 1.0, p < .001 for all cases\footnote{Full summary statistics reported in the supplementary materials at https://osf.io/de5hc/.}).

The distribution of recommended sharing percentages is shown in Figure 4. Of those who chose a sharing percentage, very few chose 0\% sharing, which indicates widespread support for the notion of some redistribution of the economic value of personal data. However, as with the distribution of prices, no statistical analyses were undertaken due to the sample size. 

\begin{figure}[h]
  \centering
  \includegraphics[width=\linewidth]{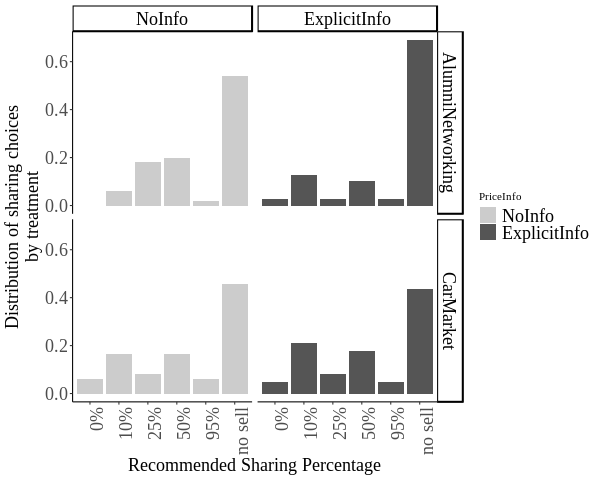}
  \caption{Distribution of sharing percentages, including refusal to select an option because data should not be sold.  
}
  \Description{Refusal to price.}
\end{figure}

\section{Discussion}

If lay people support the current legal and industry allocation of personal data via sales in commercial markets, then judged distress and the rate of refusals to price in a typical and legal data sale should be low in all cases. This is not reflected in the results. 

An alternative theory is that perhaps lay people do not agree that all legal data sales are acceptable but rather apply the same cultural rules that \cite{McGraw05} highlighted, notably that some relational framings make personal data sales acceptable while others do not. If personal data are like market goods, relational framing should have substantial effects on judgments of a sale, such that a pure market framing makes a transaction acceptable to participants, as concretely manifested via low or zero rates of refusing to price. This is also not reflected in the results.     

Refusal to price is not due to uncertainty about market price, which is controlled for in the explicit information condition. Refusal to price is similarly not due to disinterest in the price information, which is controlled for in the veiled information condition. People can recognize that personal data has a market value and be interested to know the price, but still condemn the idea of such a market and also refuse to participate when they have the option of refusal. The experimental results support various proposals that would move the emphasis on privacy law away from a market-premised, contractual model of privacy based on notice and consent \cite{Strahilevitz10} \cite{Davis19} \cite{Richards16}.

The results also speak to existing scholarship on data labor and personal data as property. Some scholars and politicians alike have proposed that popular discontent with the big data society and commercial data collection might, at least partly, be addressed by redistribution of the economic gains of data collection through the recognition and financial compensation of data ownership or data labor \cite{Fisher19} \cite{Posner18}. The results of this study, specifically that the option to share economic proceeds does not reduce rejection of data sales, suggest that data property and revenue sharing options would have to be carefully designed to be consistent with the conceptual categories ordinary people ascribe to their data. Transferring or sharing the ownership of personal data and derived economic income streams may not be enough to address the concerns of ordinary people about personal data markets if such suggested reforms are perceived to merely extend or reinforce personal data markets.

In this experiment, market appropriateness was discussed as a binary indicator, but it is more likely on a continuous spectrum, with people relatively more or less accepting of market allocation rather than strictly accepting or not. The results here simply establish an important divergence between cultural norms on the one hand and the current form of typical and legal data transfers, as operationalized by a discussion of market appropriateness. There is more to investigation, and future research can better delineate acceptable modes and spheres of exchange for personal data on a spectrum that may better elucidate how ordinary people make trade-offs among culturally incommensurate goods, particularly in the domain of technology interacting with core personal values, such as those implicated by privacy.

\section{Limitations}

There are limitations intrinsic to the methodology of the study. Those who participate in online forums are more technologically knowledgeable than the general population and so may not be fully representative of the reactions of a more heterogeneous population \cite{Redmiles19}. In this experiment, however, any potential sampling concerns  tend to undercut the possibility that the experimental results occur due to the sampling venue or methodology. Quite the opposite, the participants in this study earned money by providing some form of information about themselves. Those  who do not participate in online survey markets are likely even more resistant to data markets than the instant data pool.

Also, this study does not support the broad conclusions that all people reject the notion of data markets. Indeed, some participants did agree to price the data, did then price at market rates, and did suggest low percentages for sharing the proceeds. It may be that the population has heterogeneous attitudes about data sales, and that there is a distinct population of lay people fully on board with personal data markets even when others are not. Such heterogeneity should be examined in future studies. This possibility does not negate the results or conclusion of this study but rather provides experimental evidence to support concerns raised by e.g. \cite{Taylor04} regarding the distributive effects of data markets. 

Another possible critique of the study is that it does not replicate the model vignette of \cite{McGraw05} despite taking inspiration from that study. Thus a conclusion that personal data are not like pens is not directly supported by the experimental results. But, the results of this study stand on their own. The main result of this study is that high rates of participants refuse to price data in a typical and legal market transaction. The results do not need comparison with \cite{McGraw05}.      

Another limitation relates to the relevance of the study, namely whether the lack of cultural acceptance of a market for personal data should matter from a legal perspective. Many economically significant activities are also culturally significant, even sacred, such as getting married or mourning a death. Why should the market framing of personal data be problematic if other significant events can be handled with contracts and monetary pricing, such as contracts with wedding planners or standard pricing at funeral homes? But this oversimplifies the matter. Law does recognize cultural categories as relevant to legal ones. Consider that the law for gifts is not the same as that for market transactions, and that such legal designations can impact individuals’ willingness to participate in different forms of exchange \cite{Titmuss97}.\footnote{Similar arguments can be made regarding the study's relevance to the Responsible AI movement.} Culture and law alike do consistently recognize that norms and expectations for market transactions are not the same as those in other spheres of exchange \cite{Mauss25}, and doing so in the case of personal data should not be considered particularly unusual. 

\section{Conclusion}

This study suggests that the current legal regime and market realities governing personal data may not provide conceptual consistency with the intuitions or desires of ordinary people. Lay people refuse data sales at high rates, even when there is an opportunity to share the economic gains of such sales with data subjects. Likewise, lay people refuse data sales at high rates even after voluntarily learning about the market rates for data. Lay people are thus aware of data markets as a concept and can be interested in them, all while rejecting personal data sales. The results have direct implications for privacy law and for responsible AI. Privacy law regimes premised on market metaphors will likely continue to fail in meeting the expectations and desires of ordinary people, and so legal reforms should look beyond contractual models when seeking to update privacy laws. Likewise, the responsible AI community should look beyond notice and consent paradigms to account for cultural norms when considering fair data collection and exchange practices.

\begin{acks}
Funding was provided by ETH Zurich's Center for Law and Economics. For helpful comments and insights on earlier drafts, the author thanks Svetlana Yakovleva and Alexander Stremitzer, as well as participants of the Information Law Institute Privacy Research Group at NYU Law School, the ICML 2020 Workshop on Economics of Privacy and Data Labor, and the Privacy Law Scholars’ Conference 2020.
\end{acks}

\bibliographystyle{ACM-Reference-Format}
\bibliography{sample-base}

\end{document}